\documentclass[12pt]{iopart}

\usepackage{iopams}
\usepackage{graphicx}
\usepackage{setstackmod}

\newcommand{\bra}[1]{\mbox{$\langle #1|$}}
\newcommand{\ket}[1]{\mbox{$|#1\rangle$}}

\begin{document}

\title{Matchgate quantum computing and non-local process analysis}

\author{S. Ramelow$^{1,2}$, A. Fedrizzi$^{1}$\footnote{These authors contributed equally to this work.}, A. M. Steinberg$^{1,3}$, A. G. White$^{1}$}

\address{$^1$Department of Physics and Centre for Quantum Computer
Technology, University of Queensland, QLD 4072, Australia}
\address{$^2$Faculty of Physics, University of Vienna, Boltzmanngasse 5, 1090
Vienna, Austria}
\address{$^{3}$Center for Quantum Information and Quantum Control, Institute for Optical Sciences, Department of Physics, University of Toronto, 60 St. George Street, Toronto, ON, M5S 1A7, Canada}

\eads{\mailto{sven.ramelow@univie.ac.at}, \mailto{a.fedrizzi@uq.edu.au}}

\begin{abstract}
In the circuit model, quantum computers rely on the availability of a universal quantum gate set. A particularly intriguing example is a set of two-qubit only gates: ``matchgates'', along with \textsc{swap} (the exchange of two qubits). In this paper, we show a simple decomposition of arbitrary matchgates into better known elementary gates, and implement a matchgate in a linear-optics experiment using single photons. The gate performance was fully characterized via quantum process tomography. Moreover, we represent the resulting reconstructed quantum process in a novel way, as a fidelity map in the space of all possible nonlocal two-qubit unitaries. We propose the \textit{non-local distance} --- which is independent of local imperfections like uncorrelated noise or uncompensated local rotations --- as a new diagnostic process measure for the non-local properties of the implemented gate. 
\end{abstract}

\date{\today}

\maketitle

In quantum computation, an essential requirement of the circuit model is a universal gate set, which enables the approximation of any given unitary process to arbitrary precision \cite{nielsen2000qca}. The advantages and disadvantages of various gate sets are still being actively explored; for example, one set may be more natural than others for interpreting a certain problem, or, in a given physical architecture one set may require far less resources than another.
The best known gate-set class is that of any entangling 2-qubit gate in combination with arbitrary single-qubit unitaries \cite{dodd2002uqc}: most famously the 2-qubit \textsc{cnot} gate in conjunction with the 1-qubit Hadamard, \textsc{h}, and phase, \textsc{t}, gates. Circuits constructed of solely the \textsc{cnot} and \textsc{h} gates can be simulated efficiently with a classical computer \cite{nielsen2000qca}. However, the addition of the \textsc{t} gate---itself also efficiently simulatable---enables universal quantum computing (which of course is generally believed \emph{not} to be efficiently simulatable). Another important gate-set class uses 3-qubit entanglers, such as the Toffoli gate---which has only recently been demonstrated in linear optics \cite{lanyon2008sql} and ion traps \cite{monz2008rqt}---along with \textsc{h}.

In this paper we demonstrate a new gate-set class based \emph{only} on 2-qubit gates, specifically the \emph{matchgate} \cite{valiant2002qcc}, which can be entangling, and the \textsc{swap} gate, which is strictly non-entangling \cite{terhal2002csn,jozsa2008mac}. Matchgates---originally introduced in graph theory \cite{valiant2002qcc}---are 2-qubit unitaries,
\begin{equation}
\textsc{g}_\textsc{ab}=\left(\begin{array}{cccc}
a_{11} & 0 & 0  & a_{12}  \\
0 & b_{11} & b_{12}  & 0  \\
0 & b_{21} & b_{22}  &  0 \\
a_{21} & 0 & 0 & a_{22} \\
\end{array}
\right), \label{eq:matchgate}
\end{equation}
where the 1-qubit unitaries
\begin{equation}
\textsc{a}=\left(\begin{array}{cc}
a_{11} & a_{12}\\
a_{21} & a_{22}\\
\end{array}
\right)~~\mathrm{and}~~\textsc{b}=\left(\begin{array}{cc}
b_{11} & b_{12}\\
b_{21} & b_{22}\\
\end{array}\right),
\end{equation}
are members of $U(2)$ with $\textrm{det}(\textsc{a}){=}\textrm{det}(\textsc{b})$, and act on the even and the odd 2-qubit parity subspaces respectively. Note that the determinant condition precludes gates such as \textsc{swap}, which otherwise could be constructed as $\textsc{g}_\textsc{ix}{=}\textsc{swap}$, where \textsc{i} is the 1-qubit identity. Matchgates include a rich number of entangling gates---including maximal entangling gates---as well as many classes of local gates. If matchgates act only between nearest-neighbour qubits, then the resulting circuit can be efficiently simulated classically \cite{valiant2002qcc}. In this context matchgates relate to systems of non-interacting fermions \cite{terhal2002csn}. Moreover, they are connected to 1D quantum Ising models: explicit matchgate circuits for simulations of such strongly-correlated quantum systems have been constructed in \cite{verstraete2009qcs}. However, if matchgates are also allowed to operate between \emph{next}-nearest neighbours---a seemingly trivial change achieved via \textsc{swap} gates---then the circuit can perform universal quantum computation \cite{jozsa2008mac}. Clearly, the resulting universal gate set is entirely different from the two mentioned above.

Here we show how to realise arbitrary matchgates using circuit elements already demonstrated in a range of physical architectures; we go on to demonstrate and measure matchgate operation in a linear optics photonic system, quantifying their performance with a new method for experimental analysis of two-qubit gates.

\begin{figure}[!thbp]
\begin{center}
\includegraphics[width=0.6\textwidth]{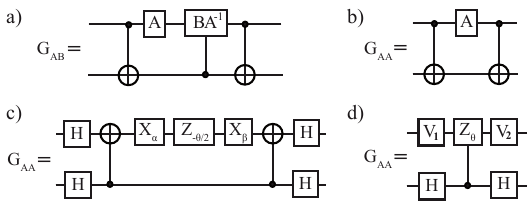}
\caption{a) Decomposition of general matchgates $\textsc{g}_\textsc{ab}$. b) Simplified decomposition of symmetric matchgates $\textsc{g}_\textsc{aa}$ as described in the text. c) The same gate after flipping the \textsc{cnot}s with $\textsc{h}$s and decomposing $\textsc{h} \textsc{a} \textsc{h}$ into $\textsc{x}$- and $\textsc{z}$-rotations. d) Further simplification shows that $\textsc{g}_\textsc{aa}$, up to a global phase, can be implemented using a single \textsc{cz}$_{\theta}$ and local unitaries $\textsc{v}_1{=}\textsc{hx}_{\alpha}$ and $\textsc{v}_2{=}\textsc{z}_{-\theta/2}\textsc{x}_{\beta}\textsc{h}$ where $\alpha$, $\beta$, and $\theta$ are related to $\textsc{a}$ via $\textsc{h} \textsc{a} \textsc{h}{=}\textsc{x}_{\alpha}\textsc{z}_{-\theta/2}\textsc{x}_{\beta}$. For $\textsc{g}_\textsc{hh}$, $\theta{=}\pi$, and $\textsc{cz}_{\theta}\rightarrow\textsc{cz}$.}
\label{fig:decomp}
\end{center}
\end{figure}

Fig.~\ref{fig:decomp}a) shows the decompostion of an arbitrary matchgate, $\textsc{g}_\textsc{ab}$, into \textsc{cnot} ($\textsc{cnot}{=}\ket{0}\bra{0}\otimes\textsc{i}{+}\ket{1}\bra{1}\otimes\textsc{x}$), controlled-unitary and single-qubit gates. Recall that, depending on the parity of the input state into the matchgate, either $\textsc{a}$ or $\textsc{b}$ acts on the two qubits. The first step of our general decomposition is to encode the parity of the two-qubit state onto one of the qubits. The first \textsc{cnot} gate turns the bottom qubit into state $\ket{0}$ for even-parity inputs ($\ket{00} {\rightarrow} \ket{00}$ \& $\ket{11} {\rightarrow} \ket{10}$) and into $\ket{1}$ for odd-parity inputs ($\ket{01} {\rightarrow} \ket{01}$ \& $\ket{10}  {\rightarrow} \ket{11}$). This qubit then acts as the control for the controlled unitary, \textsc{cu} ($\textsc{cu}{=}\ket{0}\bra{0}\otimes\textsc{i}{+}\ket{1}\bra{1}\otimes\textsc{u}$), where $\textsc{u} {=} \textsc{b} \textsc{a}^{-1}$. If the bottom qubit is in state $\ket{0}$ (parity${=}0$), \textsc{a} will act on the top qubit; if it is in state $\ket{1}$ (parity${=}1$), \textsc{cu} will undo \textsc{a} before it performs \textsc{b}. The final \textsc{cnot} gate returns the qubits from the parity encoding to the original basis.

Note that, alternatively, any 2-qubit operation could be implemented with three \textsc{cnot} gates and 8 single-qubit unitaries \cite{vidal2004uqc}, or two ``$\mathcal{B}$''-gates (yet to be experimentally demonstrated) and 6 single-qubit unitaries \cite{zhang2004mct}. Our decomposition, which offers a starting point for further simplifications, in contrast requires fewer gates and allows one to recast matchgate circuits (e.g. those in \cite{verstraete2009qcs}) into circuits consisting of more familiar quantum gates ---the \textsc{cnot} \cite{obrien2003dao,gasparoni2004rpc}, the more general \textsc{cu} \cite{lanyon2008sql}, and single qubit rotations, all of which have been individually implemented in various architectures.

We now show how to build the matchgates required for a universal gate set. The universality proof in \cite{jozsa2008mac}, relies on showing how matchgates and \textsc{swap} can be applied to implement another universal set---\textsc{h}, \textsc{t} and \textsc{cz}. As outlined in the Appendix, each logical qubit has to be encoded in two (or four) physical qubits. The required matchgate set is then $\textsc{g}_\textsc{xx}$, $\textsc{g}_\textsc{tt}$ and $\textsc{g}_\textsc{hh}$. These gates are all symmetric i.e. $\textsc{a}{=}\textsc{b}$, and the circuit of Fig.~\ref{fig:decomp}a) is greatly simplified, because $\textsc{a}^{-1}\textsc{a}{=}\textsc{i}$: the controlled unitary turns into a ``controlled identity''. The resulting circuit diagram, shown in Fig.~\ref{fig:decomp}b), still requires two 2-qubit gates. It does however resemble the general construction of a \textsc{cu} \cite{barenco1995egq} gate, i.e. it can be replaced by a single \textsc{cu} and 4 single-qubit unitaries in the following way: We flip the \textsc{cnot}s upside down by adding 4 Hadamards---$\overline{\textsc{cnot}}{\equiv}(\textsc{h}{\otimes}\textsc{h}){\times}\textsc{cnot}{\times}(\textsc{h}{\otimes}\textsc{h})$---and rewrite the resulting central unitary $\textsc{h} \textsc{a} \textsc{h}$, up to a global phase, as $\textsc{x}_{\alpha}\textsc{z}_{-\theta/2}\textsc{x}_{\beta}$ \cite{barenco1995egq}, Fig.~\ref{fig:decomp}c). The rotations $\textsc{x}_{\alpha}$ and $\textsc{x}_{\beta}$ commute with their respective \textsc{cnot}s, which allows us to express the two-qubit part of the operation as a single controlled $\textsc{cz}_{\theta}$, Fig.~\ref{fig:decomp}d). This simplified circuit for $\textsc{g}_\textsc{aa}$ can now be directly implemented in bosonic systems using the technique of shortcuts through higher-dimensional Hilbert spaces \cite{lanyon2008sql}.

\begin{figure}[h!]
\begin{center}
\includegraphics[width=0.7\textwidth]{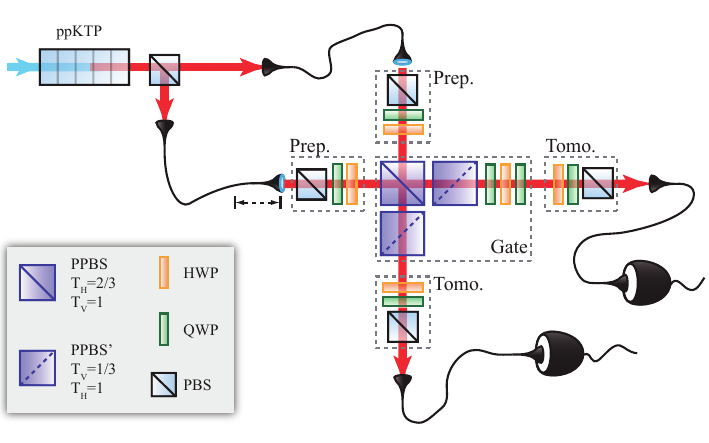}
\caption{Experimental scheme. Orthogonally polarized photon pairs are created in a nonlinear ppKTP crystal which is pumped by a 410 nm laser diode,  using focussing parameters from \cite{fedrizzi2007wtf}. The photons are split at a polarizing beamsplitter (PBS), collected, and guided to the circuit with single-mode optical fibers. Their input polarizations are set with a PBS, one quarter- and one half-wave plate (QWP, HWP) (Prep.). They are then superposed on a partially polarizing beamsplitter (PPBS) which transmits $2/3$ of horizontal, $\ket{H}$ and perfectly reflects vertical, $\ket{V}$ light. Loss elements (PPBS$^{\prime}$) correct the respective amplitudes and the unknown phase shift at the central PPBS is compensated by a combination of a QWP, a HWP and another QWP in one output port of the PPBS. If compensated correctly, only the input term $\ket{HH}$ picks up a phase shift of $\pi$ due to non-classical interference, which realizes a bit-flipped \textsc{cz} when measured in coincidence. The photons are then jointly analyzed with a QWP, a HWP and a PBS (Tomo.) before they are detected by two single-photon detectors. The rotations for the 4 single-qubit unitaries required for $\textsc{g}_{\textsc{hh}}$ in Fig.~\ref{fig:decomp}d) are incorporated into the preparation and measurement waveplate settings.}
\label{fig:scheme}
\end{center}
\end{figure}

Turning back to the universal matchgate set, we find that $\textsc{g}_\textsc{xx}{=}\textsc{x}{\otimes}\textsc{x}$ and $\textsc{g}_\textsc{tt}{=}\textsc{t}{\otimes}\textsc{i}$---both operations are local and can be done trivially in a photonic architecture with waveplates or interferometers. Similarly, \textsc{swap} is a straightforward procedure in optics, either in free-space or integrated circuits. The one non-trivial gate to be demonstrated is
\begin{equation}
\textsc{g}_\textsc{hh}{=}\frac{1}{\sqrt{2}}\left(\begin{array}{cccc}
1 & 0 & 0  & 1  \\
0 & 1 & 1  & 0  \\
0 & 1 & -1  &  0 \\
1 & 0 & 0 & -1 \\
\end{array}\right),\label{eq:ghh}
\end{equation}
which is a nonlocal, maximally entangling gate. For this gate the decomposition yields $\theta{=}\pi$ and therefore $\textsc{cz}_{\theta}$ in Fig.~\ref{fig:decomp}d) is the well known \textsc{cz} gate. Experimentally, we can therefore implement $\textsc{g}_\textsc{hh}$ using polarization-encoded single photons, partially polarizing beam splitters and coincidence detection \cite{langford2005dse,kiesel2005loc,okamoto2005doq}. The experimental setup is explained in Fig.~\ref{fig:scheme}.

\begin{figure}[h!]
\begin{center}
\includegraphics[width=0.9\textwidth]{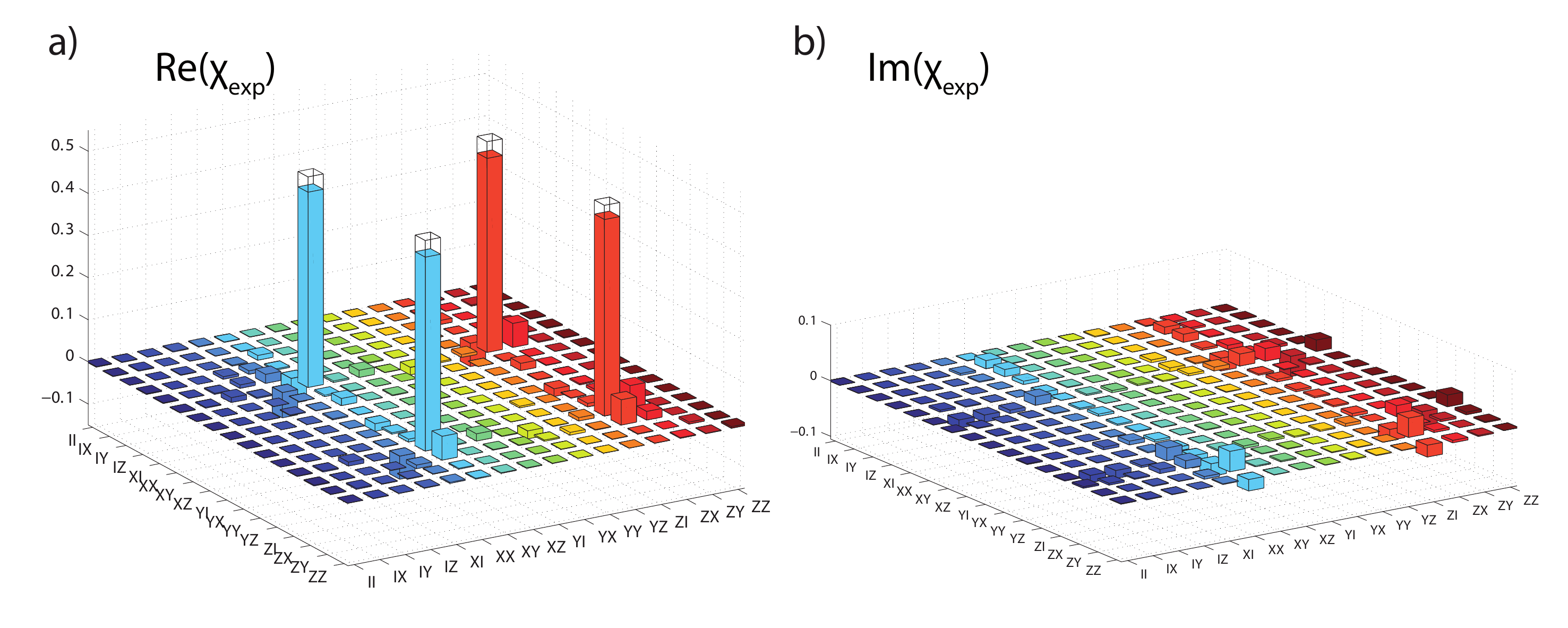}
\caption{a) Real and b) imaginary parts of the $\textsc{g}_\textsc{hh}$ process matrix $\chi_{exp}$ reconstructed from quantum process tomography measurements. The wireframes represent the ideal process $\chi_{ideal}$. The process fidelity of $\chi_{exp}$ with $\chi_{ideal}$ was $92.3\pm0.2\%$. We did not apply any numeric local rotations to optimize this result in contrast to, e.g. \cite{lanyon2008sql}.}
\label{fig:chi}
\end{center}
\end{figure}

We characterized our gate using full quantum \emph{process} tomography \cite{obrien2004qpt}, preparing $16$ combinations of the states $\{\ket{H},\ket{V},\ket{D}{=}1/\sqrt{2}(\ket{H}{+}\ket{V}),\ket{R}{=}1/\sqrt{2}(\ket{H}{+}i\ket{V})\}$ at each input and projecting into an overcomplete set of 36 measurements at the outputs. The photons were then detected in coincidence. The resulting process matrix $\chi_{exp}$, reconstructed via maximum-likelihood estimation, is shown in Fig.~\ref{fig:chi}. The process fidelity \cite{gilchrist2005dmc} with $\chi_{ideal}$ is $92.3\pm0.2\%$, where the error was calculated assuming Poissonian count statistics. We attribute the remaining errors to non-ideal waveplates, imperfections in the spatial and temporal mode-overlap, as well as in the splitting ratios of the PPBS's.

The question remains: how well we have actually implemented a matchgate? A single process fidelity, such as calculated above, reveals the overlap between the experimental process and the corresponding, ideal target process. However, it does not yield any information about the way in which the process is not ideal---is it just mixed due to random noise or are we in fact implementing a different unitary process than we actually thought? In particular, we are interested in the \emph{nonlocal} properties of the quantum process, as they define its entangling power and errors in them cannot be corrected with local operations.

Interestingly, as shown in \cite{kraus2001oce}, out of the $15$ real parameters which define a unitary 2-qubit operator $U{\in}SU(4)$, only three actually describe the nonlocal part of the unitary, the remaining $12$ relating to local transformations:
\begin{equation}
U{=}(u_{1}{\otimes}v_{1})e^{-\frac{i}{2}(c_{1}\sigma_{x}\otimes\sigma_{x}+c_{2}\sigma_{y}\otimes\sigma_{y}+c_{3}\sigma_{z}\otimes\sigma_{z})}(u_{2}{\otimes}v_{2}).\label{eq:canonical}
\end{equation}

\begin{figure}[!t]
 \begin{center}
 \includegraphics[width=0.94\textwidth]{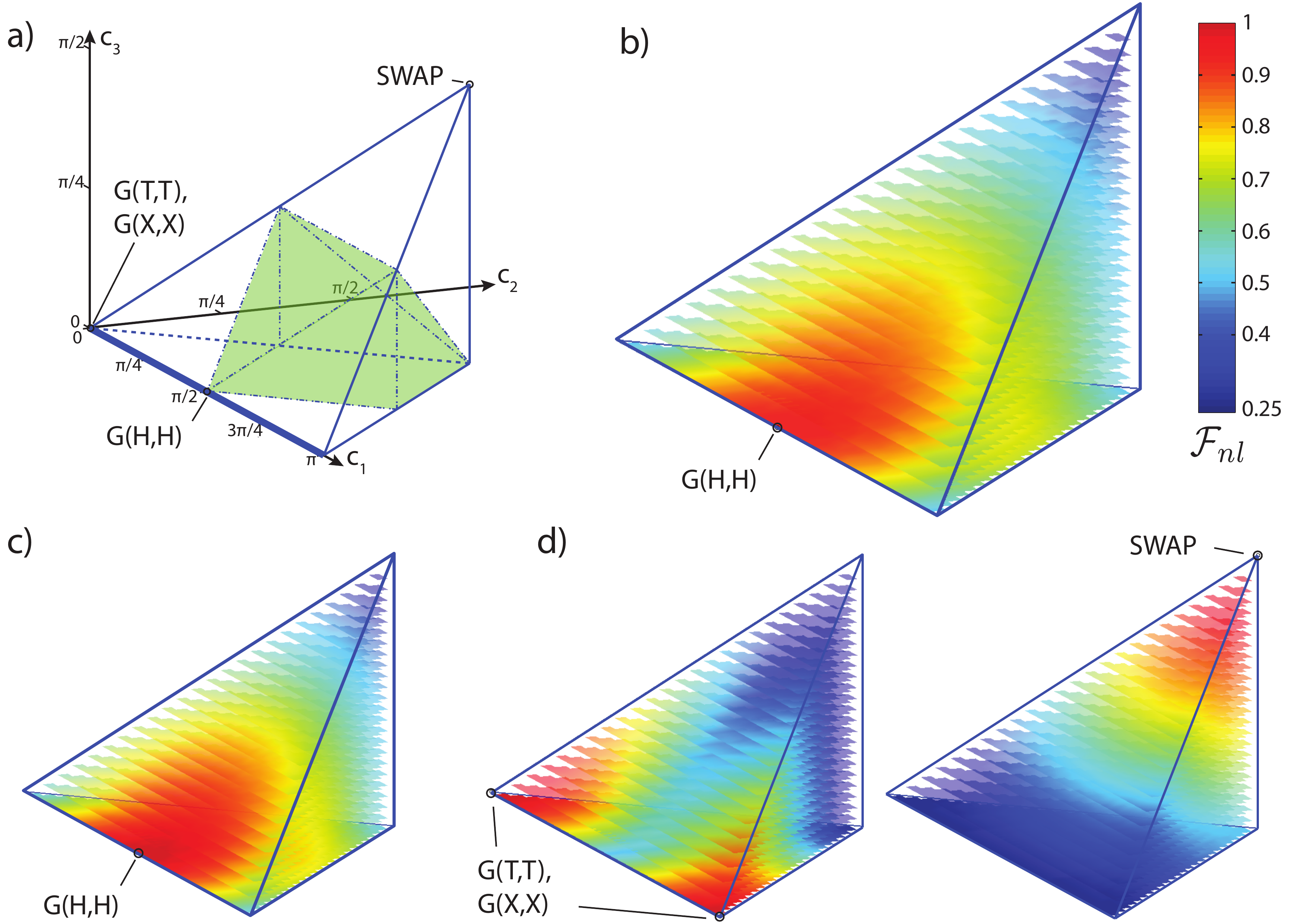}
 \caption{a) Tetrahedron depicting the space of all nonlocal two-qubit unitary gates. All local gates $u{\otimes}v$, like $\textsc{g}_{\textsc{xx}}$ and $\textsc{g}_{\textsc{tt}}$, are represented by $[0,0,0]$; \textsc{swap} is located at $[0,\pi/2,\pi/2]$. The shaded volume represents the space of all perfect entanglers \cite{zhang2003gtn}, i.e. gates that can turn separable states into Bell states. The (thick blue) line $[\gamma,0,0]$ contains all $\textsc{g}_{\textsc{aa}}$ gates, which are equivalent to \textsc{cu} \cite{lanyon2008sql}. $\textsc{g}_{\textsc{hh}}$ is located at its midpoint $[\pi/2,0,0]$, and is therefore the only symmetric matchgate which is maximally entangling. b) Fidelity map for the experimentally reconstructed $\textsc{g}_{\textsc{hh}}$ process matrix showing the fidelity (Eq.~\ref{eq:fidelity}) with respect to 6201 evenly spread ideal 2-qubit operators and optimising over all local unitary degrees of freedom. The maximum fidelity is $94.7\pm0.3\%$ at ${\sim}[\pi/2,0,0]$ which is exactly the point representing $\textsc{g}_{\textsc{hh}}$. c-d) Theoretical fidelity maps for several ideal gates: the target gate $\textsc{g}_{\textsc{hh}}$ and the local gates $\textsc{g}_{\textsc{xx}}$ and $\textsc{g}_{\textsc{tt}}$ as well as \textsc{swap}.}
\label{fig:weyl}
\end{center}
\end{figure}

This decomposition allows a very intuitive geometrical representation of local equivalence classes. The three nonlocal parameters $c_{1}$, $c_{2}$ and $c_{3}$ can be used to construct a 3-dimensional space of nonlocal gates. Symmetries reduce this space to a tetrahedron, shown in Fig.~\ref{fig:weyl} (a), the so-called Weyl-chamber \cite{zhang2003gtn}. Each point in this chamber represents all locally-equivalent gates with unique nonlocal properties (with the exception of the $c_{3}$=0 plane, which is symmetric around the line $[\pi/2,0,0]\rightarrow[\pi/2,\pi/2,0]$); the \textsc{cnot} gate, \textsc{cz}, and therefore also $\textsc{g}_\textsc{hh}$ gate, are all locally equivalent and located at $[\pi/2, 0, 0]$. Using a method from \cite{zhang2003gtn}, one can directly check whether two gates are locally equivalent, which confirms the results obtained from our matchgate decomposition in Fig.~\ref{fig:decomp}. 

In order to illustrate our experimental process in the Weyl chamber and to find the nonlocal unitary gate closest to it, we numerically translated $6201$ evenly-spaced ideal 2-qubit operators defined by (\ref{eq:canonical}) into their process representation and then calculated their maximal \emph{nonlocal} process fidelity,
\begin{equation}
\mathcal{F}_{nl}(c_{1},c_{2},c_{3})=\underset{u_{1}...v_{2}}{\textrm{max}}\mathcal{F}(\chi_{exp},\chi(c_{1}...c_{3},u_{1}...v_{2})).\label{eq:fidelity}
\end{equation}
to $\chi_{exp}$ by numerical optimization over the local transformations $u_{1}, v_{1}$, and $u_{2}, v_{2}$. Applying numeric local rotations to optimize a process fidelity to a specific (ideal) target process is a common practice in linear optics quantum experiments, see e.g. \cite{langford2005dse,resch_entanglement_2007,lanyon2008sql}. This is motivated by the high precision and ease with which local unitaries can be implemented in optics. We extend on this by calculating the non-local fidelity for every point in the Weyl chamber. The result is a three-dimensional process fidelity map, shown in figure~\ref{fig:weyl}b). After optimization over the local unitaries, we find a maximum process fidelity for our experimental gate of $94.7{\pm}0.3\%$ (increased from 92.3\%, figure~\ref{fig:chi}) at $[c_{1},c_{2},c_{3}]{\sim}[\pi/2,0,0]$. 

For comparison, we show the fidelity maps for the ideal $\textsc{g}_\textsc{hh}$ in figure ~\ref{fig:weyl}c). The maximum fidelity is, as expected, $1$ at $[\pi/2, 0, 0]$. The fidelity of $\textsc{g}_\textsc{hh}$ is $50\%$ with local gates and $25\%$ with \textsc{swap}, which reflects the fact that the distance in the Weyl chamber to the latter is maximal ($\mathcal{F}$ has a range of $\mathcal{F}\in[0.25,1]$). The volume of nonlocal gates with $\geq$90\% fidelity to $\chi_{ideal}$ is 11.6\%. For $\chi_{exp}$ this volume shrinks to 4.85\%, due to decoherence. Ideal fidelity maps for the remaining gates from the universal matchgate set, $\textsc{g}_\textsc{xx}$, $\textsc{g}_\textsc{tt}$ and $\textsc{swap}$ in figure~\ref{fig:weyl}d) complete the picture.

We can now define a new process distance measure---the \emph{nonlocal distance}, $\Delta_{nl}{\equiv} \sqrt{\Delta c_{1}^{2}+\Delta c_{2}^{2}+\Delta c_{2}^{2}}$, where $\Delta_{nl}{\in}[0,\pi]$ and $\Delta c_{i}$ is the difference of the coordinates of a target unitary gate and the coordinates for the maximum fidelity for a given process $\mathcal{F}_{nl}(c_{1},c_{2},c_{3})_{\max}$ obtained from the optimization, Eq.~(\ref{eq:fidelity}). This distance, can of course also be used for two ideal unitaries in the chamber. According to \cite{gilchrist2005dmc}, $\Delta_{nl}$ meets all distance measure criteria for pure processes and can be seen as a \emph{diagnostic measure}. To illustrate this let us compare the process purity to $\mathcal{F}_{nl}$: because it is maximized to an underlying, unitary 2-qubit operation, stripped of its local rotations, it will signal---similar to the purity---when an implemented process is pure. However, in contrast to the purity, $\mathcal{F}_{nl}$ also tells us which operation was in fact implemented.

Our experimental gate is located at a distance $\Delta_{nl}{\sim}0$ from the ideal \textsc{cz}. The implemented gate therefore has---within negligible uncertainties dominated by the numerical optimization---exactly the intended non-local properties. Upon closer inspection, this is a direct consequence of the way the gate is implemented physically. First of all, any uncorrelated noise process, such as depolarization or dephasing \cite{nielsen2000qca}, acting on the input or output states of our gate is non-unitary and adds mixture to the process, which leads to a uniform decrease of $\mathcal{F}_{nl}$ over the whole non-local space, as illustrated by comparison of the experimental to the ideal fidelity map in figure \ref{fig:weyl} b) and c). Second, imperfections in optical components such as the PPBS central to our gate equally lead to mixing because the underlying 2-qubit operations for reflectivity values close to $\eta_{ideal}{=}1/3$ are non-unitary \cite{ralph2002loc}, and do not result in a shift of the gate's position in the unitary space represented by the Weyl chamber. Other imperfections, such as temporal or spatial mode mismatch, can be modelled as incoherent sums of different unitary operations. In this case, fidelity maps would still not show a shift of the non-local location of the resulting process, but rather the emergence of additional local maxima, centred at the non-local positions of the contributing unitary operators. Our optical setup therefore does not have any \emph{non-local, unitary} degrees of freedom in the sense that the available experimental parameters do not allow an actual movement of the fidelity maximum in the Weyl chamber. The physical implementation of the $\textsc{cu}$ gates demonstrated in \cite{lanyon2008sql}, in contrast, would have allowed a shift of the experimental process along the line marking the symmetric matchgates. 

Our non-local process analysis allows us to conclude that the overall quality of our gate implementation cannot be improved by unitary corrections beyond simple local rotations. In other words, apart from local rotations which can be easily applied, the closest unitary gate to the experimentally implemented process is indeed $\textsc{g}_{\textsc{hh}}$. The remaining reduction in fidelity is identified as mixture, caused by effects such as higher-order photon emissions from the SPDC source, which is supported by the measured, non-ideal process purity of $89.8\pm0.4\%$. Future experimental work will have to focus on removing these noise sources.

In summary, we have implemented a matchgate which allows universal computation when combined with the simple two-qubit \textsc{swap}. Our gate decomposition provides a simple procedure to implement known matchgate circuits in linear optics, and, vice versa, translate quantum computing algorithms formulated in terms of more common gate sets into physical architectures which are naturally suited to implement matchgates. 

We characterized our gate using quantum process tomography and illustrated the fidelity overlap of the experimental process with all possible nonlocal gates in the Weyl chamber. We expect this method to develop into a valuable diagnostic tool in quantum information processing, especially in the analysis of noisy processes where it can help identify and distinguish unitary error sources from genuine mixing. Suggested lines for further research in this topic are, for example, how correlated noise influences the nonlocal properties of a quantum gate and nonlocal process discrimination \cite{laing2009eqp}. 

\section*{Acknowledgments}
We wish to thank B.~P.~Lanyon and N.~K.~Langford for valuable input. We acknowledge financial support from the Australian Research Council Discovery and Federation Fellow programs and an IARPA-funded U.S. Army Research Office contract. SR acknowledges financial support by the FWF project CoQuS No. W1210-N16. AMS acknowledges support by the Natural Sciences and Engineering Research Council of Canada, Quantum Works and the Canadian Institute for Advanced Research.

\appendix
\setcounter{section}{1}
\section*{Appendix A} 
\begin{figure}[htbp]
\begin{center}
\includegraphics[width=0.6\textwidth]{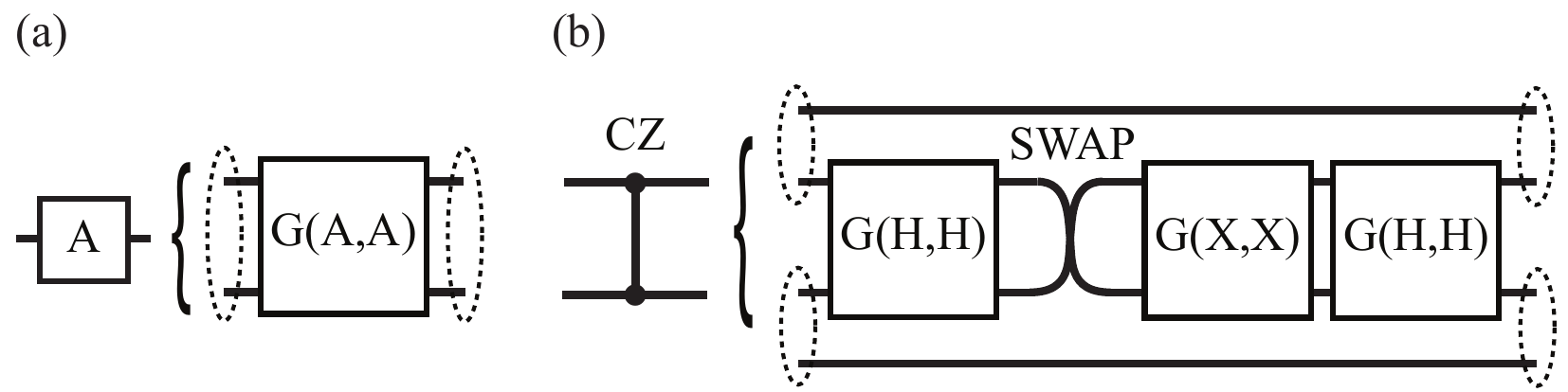}
\caption{Matchgate universality proof \cite{jozsa2008mac}. a) Symmetric matchgates $\textsc{g}_{\textsc{aa}}$ act on two physical qubits which represent one logical qubit. b) \textsc{cz} can be constructed using symmetric matchgates and \textsc{swap}. Dotted ellipses represent logical qubits.}
\label{fig:encoding}
\end{center}
\end{figure}

We briefly review the universality proof given by R. Jozsa and A. Miyake in \cite{jozsa2008mac}, which involves showing how one can construct an already known universal set from $\textsc{g}_\textsc{ab} {+} \textsc{swap}$. Because $\textsc{g}_\textsc{ab}$ is intrinsically a two-qubit gate and does not allow a change of parity of the input state, it is not straightforward to construct arbitrary single-qubit unitaries. This is solved in \cite{jozsa2008mac} by encoding the logical qubits $\ket{0}_{L}$, $\ket{1}_{L}$, into two physical qubits which have the same parity $\ket{0}_{L}{=}\ket{00}, \ket{1}_{L}{=}\ket{11}$. A symmetric $\textsc{g}_\textsc{aa}$ acting on the two physical qubits  then performs the single-qubit unitary $\textsc{a}$ on the logical qubit, Fig.~\ref{fig:encoding}a). This 2-qubit encoding is universal with next-next-nearest neighbor interactions, whereas a more complicated 4-qubit encoding, which requires the non-trivial $\textsc{g}_\textsc{zx}$, allows universal quantum computing with only next-nearest neigbour interactions \cite{jozsa2008mac}. 

Next, we need an entangling gate which can act between two encoded logical qubits. This can be achieved by adding \textsc{swap}, because, as illustrated in Fig.~\ref{fig:encoding}b), $\textsc{cz}{=} \textsc{g}_\textsc{hh} {\times}\textsc{swap} {\times} \textsc{g}_\textsc{xx} {\times} \textsc{g}_\textsc{hh}$, where $\textsc{x}$ denotes the $\sigma_x$ Pauli operator. In summary, the minimal set of matchgates needed for universal quantum computing is $\textsc{g}_\textsc{hh}$ and $\textsc{g}_\textsc{tt}$, which implement the single-qubit gates $\textsc{h}$ and $\textsc{t}$ on the encoded dual-qubit space, and $\textsc{g}_\textsc{xx}$, $\textsc{g}_\textsc{hh}$ and \textsc{swap}, which together form a \textsc{cz}.

\section*{References}

\begin{thebibliography}{10}

\bibitem{nielsen2000qca}
M. A. Nielsen and I.~Chuang.
\newblock {\em Quantum computation and quantum information}.
\newblock Cambridge University Press, 2000.

\bibitem{dodd2002uqc}
J.~L. Dodd, M.~A. Nielsen, M.~J. Bremner, and R.~T. Thew.
\newblock Universal quantum computation and simulation using any entangling
  hamiltonian and local unitaries.
\newblock {\em Phys. Rev. A}, 65(4):040301, Apr 2002.

\bibitem{lanyon2008sql}
B. P. Lanyon, M.~Barbieri, M. P. Almeida, T.~Jennewein, T. C. Ralph, K. J. Resch,
  G. J. Pryde, J. L. O'Brien, A.~Gilchrist, and A. G. White.
\newblock {Simplifying quantum logic using higher-dimensional Hilbert spaces}.
\newblock {\em Nature Physics}, 5(2):134--140, 2008.

\bibitem{monz2008rqt}
T.~Monz, K.~Kim, W.~H\"{a}nsel, M.~Riebe, A.~S. Villar, P.~Schindler,
  M.~Chwalla, M.~Hennrich, and R.~Blatt.
\newblock Realization of the quantum toffoli gate with trapped ions.
\newblock {\em Physical Review Letters}, 102(4):040501, 2009.

\bibitem{valiant2002qcc}
L. G. Valiant.
\newblock {Quantum circuits that can be simulated classically in polynomial
  time}.
\newblock {\em SIAM Journal on Computing}, 31(4):1229--1254, 2002.

\bibitem{terhal2002csn}
B. M. Terhal and D. P. DiVincenzo.
\newblock {Classical simulation of noninteracting-fermion quantum circuits}.
\newblock {\em Physical Review A}, 65(3):32325, 2002.

\bibitem{jozsa2008mac}
R.~Jozsa and A.~Miyake.
\newblock {Matchgates and classical simulation of quantum circuits}.
\newblock {\em Proceedings of the Royal Society A: Mathematical, Physical and
  Engineering Science}, 464(2100):3089, 2008.

\bibitem{verstraete2009qcs}
F.~Verstraete, J.~I. Cirac, and J.~I. Latorre.
\newblock Quantum circuits for strongly correlated quantum systems.
\newblock {\em Physical Review A}, 79(3):032316, 2009.

\bibitem{vidal2004uqc}
G.~Vidal and CM~Dawson.
\newblock {Universal quantum circuit for two-qubit transformations with three
  controlled-NOT gates}.
\newblock {\em Physical Review A}, 69(1), 2004.

\bibitem{zhang2004mct}
J.~Zhang, J.~Vala, S.~Sastry, and K.B. Whaley.
\newblock {Minimum construction of two-qubit quantum operations}.
\newblock {\em Physical Review Letters}, 93(2):8514, 2004.

\bibitem{obrien2003dao}
J.~L. O'Brien, G.~J. Pryde, A.~G. White, T.~C. Ralph, and D.~Branning.
\newblock {Demonstration of an all-optical quantum controlled-NOT gate}.
\newblock {\em Nature}, 426(6964):264--267, 2003.

\bibitem{gasparoni2004rpc}
S.~Gasparoni, J.~W.~Pan, P.~Walther, T.~Rudolph, and A.~Zeilinger.
\newblock {Realization of a Photonic Controlled-NOT Gate Sufficient for Quantum
  Computation}.
\newblock {\em Physical Review Letters}, 93(2):20504, 2004.

\bibitem{barenco1995egq}
A.~Barenco, C.~H.~Bennett, R.~Cleve, D.~P.~DiVincenzo, N.~Margolus, P.~Shor,
  T.~Sleator, J.~A.~Smolin, and H.~Weinfurter.
\newblock {Elementary gates for quantum computation}.
\newblock {\em Physical Review A}, 52(5):3457--3467, 1995.

\bibitem{langford2005dse}
N.~K. Langford, T.~J. Weinhold, R.~Prevedel, K.~J. Resch, A.~Gilchrist, J.~L.
  O'Brien, G.~J. Pryde, and A.~G. White.
\newblock {Demonstration of a Simple Entangling Optical Gate and Its Use in
  Bell-State Analysis}.
\newblock {\em Physical Review Letters}, 95:21, 2005.

\bibitem{kiesel2005loc}
N.~Kiesel, C.~Schmid, U.~Weber, R.~Ursin, and H.~Weinfurter.
\newblock Linear optics controlled-phase gate made simple.
\newblock {\em Phys. Rev. Lett.}, 95(21):210505, Nov 2005.

\bibitem{okamoto2005doq}
R.~Okamoto, H.~F. Hofmann, S.~Takeuchi, and K.~Sasaki.
\newblock Demonstration of an optical quantum controlled-not gate without path
  interference.
\newblock {\em Phys. Rev. Lett.}, 95(21):210506, Nov 2005.

\bibitem{fedrizzi2007wtf}
A.~Fedrizzi, T.~Herbst, A.~Poppe, T.~Jennewein, and A.~Zeilinger.
\newblock A wavelength-tunable fiber-coupled source of narrowband entangled
  photons.
\newblock {\em Opt. Express}, 15(23):15377--15386, 2007.

\bibitem{obrien2004qpt}
J.~L.~O'Brien, G.~J.~Pryde, A.~Gilchrist, D.~F.~V.~James, N.~K.~Langford, T.~C.~Ralph, and
  A.~G.~White.
\newblock {Quantum Process Tomography of a Controlled-not Gate}.
\newblock {\em Phys Rev Lett}, 93:080502, 2004.

\bibitem{kraus2001oce}
B.~Kraus and J.~I. Cirac.
\newblock {Optimal creation of entanglement using a two-qubit gate}.
\newblock {\em Physical Review A}, 63(6), 2001.

\bibitem{zhang2003gtn}
J.~Zhang, J.~Vala, S.~Sastry, and K.~B.~Whaley.
\newblock {Geometric theory of nonlocal two-qubit operations}.
\newblock {\em Physical Review A}, 67:042313, 2003.

\bibitem{resch_entanglement_2007}
K. J. Resch, J. L. {O'Brien}, T. J. Weinhold, K. Sanaka, B. P. Lanyon, N. K. Langford, A. G. White.
\newblock {Entanglement Generation by {Fock-State} Filtration}.
\newblock {\em Physical Review Letters}, 98(20):203602, 2007.

\bibitem{gilchrist2005dmc}
A.~Gilchrist, N.~K.~Langford, and M.~A. Nielsen.
\newblock Distance measures to compare real and ideal quantum processes.
\newblock {\em Phys. Rev. A}, 71(6):062310, Jun 2005.

\bibitem{ralph2002loc}
T.~C. Ralph, N.~K. Langford, T.~B. Bell, and A.~G. White.
\newblock Linear optical controlled-not gate in the coincidence basis.
\newblock {\em Phys. Rev. A}, 65(6):062324, Jun 2002.

\bibitem{laing2009eqp}
A.~Laing, T.~Rudolph, and J.~L.~O'Brien.
\newblock {Experimental Quantum Process Discrimination}.
\newblock {\em Physical Review Letters}, 102(16):160502, 2009.



\end{thebibliography}

\end{document}